\documentclass[conference]{IEEEtran}
\IEEEoverridecommandlockouts
\usepackage{cite}
\usepackage{amsmath,amssymb,amsfonts}
\usepackage{algorithmic}
\usepackage{graphicx}
\usepackage{textcomp}
\usepackage{amsmath}
\usepackage{amsthm}
\usepackage{amssymb}
\usepackage{bm}
\usepackage{xspace}
\usepackage{xcolor}
\usepackage{graphicx}
\usepackage{url}
\usepackage{framed}
\usepackage{float}
\usepackage{rotating}
\usepackage{verbatim}
\usepackage{listings}
\usepackage{lscape}

\usepackage{pgfplots}
\usepackage{pgf}
\usepackage{tikz}
\usetikzlibrary{arrows,shapes.misc,chains,scopes}

\usepackage{pgfplotstable}

\usepackage{colortbl}
\usetikzlibrary{matrix}
\usetikzlibrary{calc}
\usetikzlibrary{fit}

\newcommand{\executeiffilenewer}[3]{%
	\ifnum\pdfstrcmp{\pdffilemoddate{#1}}%
	{\pdffilemoddate{#2}}>0%
	{\immediate\write18{#3}}\fi%
}

\graphicspath{{figures/}}

\newcommand{\bmm}{\begin{matrix}}
	\newcommand{\emm}{\end{matrix}}
\newcommand{\bpm}{\begin{pmatrix}}
	\newcommand{\epm}{\end{pmatrix}}

\newcommand{\bsbm}{\left[\begin{smallmatrix}}
	\newcommand{\esbm}{\end{smallmatrix}\right]}

\newcommand{\bbm}{\begin{bmatrix}}
	\newcommand{\ebm}{\end{bmatrix}}

\newcommand*{\affmark}[1][*]{\textsuperscript{#1}}

\usepackage[utf8]{inputenc}
\usepackage[english]{babel}

\usepackage{booktabs}
\usepackage{multirow}
\usepackage{siunitx}
\usepackage{smartdiagram}

\theoremstyle{definition}
\newtheorem{definition}{Definition}
\newtheorem{conjecture}{Conjecture}
\newtheorem{remark}{Remark}
\DeclareMathOperator\erfc{erfc}

\def\BibTeX{{\rm B\kern-.05em{\sc i\kern-.025em b}\kern-.08em
    T\kern-.1667em\lower.7ex\hbox{E}\kern-.125emX}}
\begin{document}

\title{Polar Code Construction for List Decoding}

\author{
	\IEEEauthorblockN{Peihong Yuan\affmark[1], Tobias Prinz\affmark[1], Georg B\"ocherer\affmark[2], Onurcan {\.I}{\c{s}}can\affmark[3], Ronald B{\"o}hnke\affmark[3], Wen Xu\affmark[3]}
	\IEEEauthorblockA{\affmark[1]\textit{Institute for Communications Engineering, Technical University of Munich} \\
		\affmark[2]\textit{Mathematical and Algorithmic Sciences Lab, Huawei Technologies France}\\
		\affmark[3]\textit{European Research Center, Huawei Technologies Duesseldorf GmbH}\\
		Email: \{peihong.yuan,tobias.prinz\}@tum.de,\\
				georg.boecherer@ieee.org,\\
				\{onurcan.iscan,ronald.boehnke,wen.dr.xu\}@huawei.com}
}

\maketitle

\begin{abstract}
A heuristic construction of polar codes for successive cancellation list (SCL) decoding with a given list size is proposed to balance the trade-off between performance measured in frame error rate (FER) and decoding complexity. Furthermore, a construction based on dynamically frozen bits with constraints among the "low weight bits" (LWB) is presented. Simulation results show that the LWB-polar codes outperform the CRC-polar codes and the eBCH-polar codes under SCL decoding.
\end{abstract}


\section{Introduction}
Polar codes were proposed in~\cite{stolte2002rekursive,arikan2009channel} and they achieve the capacity of binary input discrete memoryless channel asymptotically in the block length~\cite{arikan2009channel}. Under successive cancellation list (SCL) decoding~\cite{tal2015list}, the finite length performance of polar codes can be improved by enhancing the distance spectrum. Cyclic redundancy check (CRC)-polar codes~\cite{tal2015list} and Reed-Muller (RM)-polar codes~\cite{li2014rm} are proposed to improve the performance of polar codes with short and moderate length. Polar codes with dynamically frozen bits and in particular eBCH-polar codes are introduced in~\cite{trifonov2016polar}. A construction for multi-kernel polar codes based on the maximization of the minimum distance is proposed in~\cite{bioglio2017minimum}. The authors in~\cite{ricciutelli2017error} analyze short concatenated polar and CRC codes with interleaving and suggest careful optimization of the outer code.

In this work, we analyze methods to improve the distance spectrum for polar codes. We propose a heuristic construction to optimize the frame error rate (FER) for a given list size. We achieve this by balancing the trade-off between FER under successive cancellation (SC) decoding and maximum likelihood (ML) decoding. A "Low weight bits" (LWB) construction based on dynamically frozen bits is presented, which outperforms CRC-polar codes and eBCH-polar codes for all considered decoding list sizes.

This work is organized as follows. In Sec.~\ref{sec:pre}, polar codes are reviewed and existing methods to improve their distance spectrum are discussed. In Sec.~\ref{sec:tool}, the tool proposed in~\cite{li2012adaptive} is used to analyze the distance spectrum of polar codes. In Sec.~\ref{sec:opt}, we balance the trade-off between distance spectrum and the performance under SC decoding for a given list size. In Sec.~\ref{sec:df}, we discuss the new LWB polar code construction. We conclude in Sec.~\ref{sec:con}.

\section{Preliminaries}\label{sec:pre}
\subsection{Polar Codes}
A binary polar code of block length $n$ and dimension $k$ is defined by the polar transform $\mathbb{F}^{\otimes\log_2 n}$ and $n-k$ frozen positions, where $\mathbb{F}$ denotes the Ar\i kan kernel 
\begin{equation}
\mathbb{F}= \begin{bmatrix}
1 & 0\\
1 & 1
\end{bmatrix}
\end{equation}
and $\otimes$ denotes the Kronecker product and $(\cdot)^\otimes$ denotes the Kronecker power. 
Polar encoding can be represented by
\begin{equation}\label{eq:polartrans}
\bm{c}=\bm{u} \mathbb{F}^{\otimes\log_2 n}.
\end{equation}
The vector $\bm{c}$ is the code word. The vector $\bm{u}$ includes $k$ information bits and $n-k$ predefined frozen bits. Polar SC decoding uses the observation $\bm{y}$ and previous estimates $\hat{u}_1,\dots, \hat{u}_{i-1}$ to decode $u_i$. Both encoding and SC decoding have complexity $\mathcal{O}(n\log_2 n)$~\cite{arikan2009channel}.

The polar code construction finds the most reliable bits under SC decoding. The Monte Carlo (MC) construction was introduced in~\cite{stolte2002rekursive,arikan2009channel}, and needs extensive simulations. In this work, the Gaussian approximation~\cite{ten2004design} for density evolution~\cite{mori2009performance} with the $J$-function~\cite{dosio2016polar} and its numerical approximation~\cite{brannstrom2005convergence} are used, which has much lower complexity and performs very close to the MC construction.

To improve the coding performance, an SCL decoding algorithm was proposed in~\cite{tal2015list}. The SCL decoder provides ML-performance for polar codes if the list size $L$ is large enough and can be performed with $\mathcal{O}(Ln\log_2 n)$ complexity. For short and moderate lengths, the original polar codes with SCL decoding still perform worse than Turbo and LDPC codes because of the low minimum distance~\cite{tal2015list}. 
\subsection{CRC and Polar Code Concatenation}
The work in~\cite{tal2015list} enhance the distance spectrum of polar codes by serial concatenating an error-detecting code and a polar code. So far in literature, the distance property of CRC-polar codes can be found only through simulations.

We use CRC codes with $\ell_{\rm CRC}$ check bits as outer codes and the SCL decoder chooses the most likely codeword that satisfies the CRC. The generator polynomials are described by a hexadecimal number (Koopman Notation~\cite{koopman_crc}), e.g., '0x5b' denotes the generator polynomial $g(x)=x^7+x^5+x^4+x^2+x+1$ ($\ell_{\rm CRC}=7$).
\begin{remark}
	An interleaver between the CRC encoder and polar encoder affects the code performance significantly~\cite{ricciutelli2017error}. In this work, conventional systematic CRC encoding is used without interleaving. Instead, we optimize over the polynomial $g(x)$. No performance loss compared to interleaving is observed.
\end{remark}
\subsection{RM-polar Codes}
The second idea is called Reed-Muller (RM)-polar codes~\cite{li2014rm}. The RM-polar codes are constructed by combining the code constructions of RM codes and polar codes. Both RM and polar codes are obtained from the same polarization matrix
$
\mathbb{F}^{\otimes\log_2 n}
$.

While polar codes select information bits according to the bit reliability under SC decoding, RM codes select the information bits according to the row weight. The bits with the largest weights of their corresponding rows are selected as information bits, and the other bits are chosen as frozen bits.

The construction of RM-polar codes sacrifices some reliable bits under SC decoding in order to guarantee a better minimum distance: 
\begin{itemize}
	\item Freeze the bits with row weight smaller than a given minimum weight $w$.
	\item Choose the most reliable remaining bits as information bits.
\end{itemize}
An RM-polar code with guaranteed minimum distance ($\geq d$) can be easily constructed by using this method. With large decoding list, RM-polar codes outperform the original polar codes because of the better distance property.

However, RM-polar codes are not very flexible, because the minimum Hamming weight of RM codes has to be a power of 2. Practically, RM-polar codes do not work well for short block length. e.g., to design a $(128,64)$ RM-polar code, there are only 2 options for the minimum distance $d$: 8 (equivalent to the original polar code) or 16 (equivalent to the RM code).
\begin{remark}
	Polar codes designed for higher SNR (than the operating points) can also improve the distance property by sacrificing reliable bits under SC decoding~\cite{hussami2009performance}. 
\end{remark}

\subsection{eBCH Polar Subcodes}
The third idea is a code construction based on extended primitive binary BCH (eBCH) codes and polar codes by using dynamically frozen bits~\cite{trifonov2016polar}. Some of the frozen bits in eBCH-polar codes are so-called dynamically frozen bits, which are defined as linear combinations of previous (with smaller indices) information bits instead of predetermined values. Consider an ($n,k^\prime$) eBCH code with parity check matrix $H$  and an ($n,k$) polar code with matrix $\mathbb{F}^{\otimes\log n}$, where $k^\prime \geq k$. Let this ($n,k$) polar code be a subcode of the ($n,k^\prime$) eBCH code, i.e., 
\begin{equation}
\bm{c}H^{\rm T} \overset{(\ref{eq:polartrans})}{=} \bm{u} \mathbb{F}^{\otimes\log_2 n}H^{\rm T}=\bm{0}
\end{equation}
where $(\cdot)^{\rm T}$ denotes the transpose of a matrix. Define a constraints matrix
\begin{equation}
V = Q (\mathbb{F}^{\otimes\log_2 n}H^{\rm T})^{\rm T}.
\end{equation}
We have $\bm{u}V^{\rm T}=\bm{0}$, where the matrix $Q$ describes elementary row operations on $(\mathbb{F}^{\otimes\log_2 n}H^{\rm T})^{\rm T}$, such that all rows of $V$ end with "1" in distinct columns. The $(n-k^\prime)\times n$ matrix $V$ describes at most $n-k^\prime$ (static or dynamically) frozen bits. The position of the last "1" in every row denotes a frozen position because of SC decoding. e.g., 
\begin{equation}
V = \begin{bmatrix}
1& 0& 0& 0 & 0 & 0\\
0& 1 & 0& 0& 0 & 0\\
0& 0 & 1& 1 & 0 & 0\\
0& 1& 0& 0 & 1 & 0
\end{bmatrix}
\end{equation}
means that $u_1,u_2,u_4,u_5$ are frozen with constraints
\begin{equation}
\begin{split}
u_1 &= 0,\\
u_2 & = 0,\\
u_4 &= u_3,\\
u_5 &= u_2 = 0.
\end{split}
\end{equation}
$u_1$, $u_2$, $u_5=0$ are statically frozen bits and since $u_1, \dots, u_6$ are decoded successively, $u_3$ is unfrozen. $u_4=u_3$ is dynamically frozen and $u_6$ is unfrozen. The construction of eBCH-polar codes is as follows:
\begin{itemize}
	\item Calculate reliabilities.
	\item Freeze/dynamically freeze the bits according to $V$.
	\item Freeze more bits according to reliabilities.
\end{itemize}
Due to the property of subcodes, ($n,k,k^\prime$) eBCH-polar codes have a guaranteed distance spectrum not worse than ($n,k^\prime$) eBCH codes. $k^\prime$ is adjustable to construct a more polar-like (with better SC-performance) code or a more eBCH-like (with better ML-performance) code.

\begin{remark}
	CRC-polar codes are also a special case of polar codes with dynamically frozen bits. Consider an $\ell_{\rm CRC}$ bits CRC outer code and an ($n,k+\ell_{\rm CRC}$) polar code. At the receiver, after the list decoding of the first $k$ bits, the remaining $\ell_{\rm CRC}$ bits can be calculated just like dynamically frozen bits. Therefore, an equivalent code construction of ($n,k$) CRC-polar codes is as follows:
	\begin{itemize}
		\item[1] Construct an original ($n,k+\ell_{\rm CRC}$) polar code.
		\item[2] Dynamically freeze the last $\ell_{\rm CRC}$ information bits with the CRC rule.
	\end{itemize}
\end{remark}

\subsection{Distance Spectrum}
\theoremstyle{definition}
\begin{definition}
	For an $(n,k)$ binary linear block code the minimum distance ${\rm d}_{\rm min}$ is the minimum Hamming distance ${\rm d}_{\rm H}(\bm{c},\bm{c^\prime})$ between two distinct codewords, $\bm{c},\bm{c^\prime}$, i.e., we have
	\begin{equation}
	{\rm d}_{\rm min}(\mathcal{C}) = \min_{\substack{  \bm{c}\neq \bm{c^\prime} \\ \bm{c},\bm{c^\prime}\in \mathcal{C}}} {\rm d}_{\rm H}(\bm{c},\bm{c^\prime}) = \min_{\substack{  \bm{c}\neq \bm{0} \\ \bm{c}\in \mathcal{C}}}  {\rm w}_{\rm H}(\bm{c})
	\end{equation}
	where ${\rm w}_{\rm H}$ denote the Hamming weight of a codeword.
\end{definition}
\begin{definition}
	For an $(n,k)$ binary linear block code (with code book $\mathcal{C}$) the multiplicity of codewords with a given Hamming weight $w$ is
	\begin{equation}
	A_w=|\{ \bm{c}|\bm{c} \in \mathcal{C}, {\rm w}_{\rm H}(\bm{c})=w \}|.
	\end{equation}
\end{definition}
The distance properties of a linear block code can be described by the distance spectrum (or weight enumerator): $A_0,A_1,\dots,A_n$.\\
Given a code distance spectrum, the code performance (FER) under ML decoding can be estimated via the union bound (UB). For the binary-input AWGN channel, the UB is
\begin{equation}\label{eq:ub}
P_{\rm B}\leq P_{\rm UB}=\frac{1}{2}\sum_{w={\rm d}_{\rm min}}^{n} A_w \erfc \left( \sqrt[]{w\text{SNR}} \right).
\end{equation}
For high SNR, the UB can be well approximated by 
\begin{equation}\label{eq:aub}
P_{\rm UB} \approx \frac{1}{2}A_{\rm min}\erfc \left( \sqrt[]{{\rm d}_{\rm min}\text{SNR}} \right)
\end{equation}
where $A_{\rm min}$ denotes the multiplicity of codewords with minimum Hamming weight.
\section{Analysis of Distance Spectrum}\label{sec:tool}
In~\cite{li2012adaptive}, the authors proposed a tool to analyze the distance spectrum by using list decoding. Suppose the list contains only the codewords with the least weights if the all zero codeword is transmitted over a channel with very small noise variance. The algorithm works as follows:
\begin{itemize}
	\item[1.] Transmit the all zero codeword with very high SNR.
	\item[2.] Perform list decoding with a very large list size on the received soft information.
	\item[3.]  (optional) Delete the codewords that do not satisfy the outer code check.
	\item[4.] Find all codewords with non-zero weight in the list and the corresponding multiplicities.
	\item[5.] Calculate the approximated UB (AUB) with (\ref{eq:ub}).
\end{itemize}
We apply this method to polar codes, CRC-polar codes, RM-polar codes, eBCH-polar codes with SCL decoding.

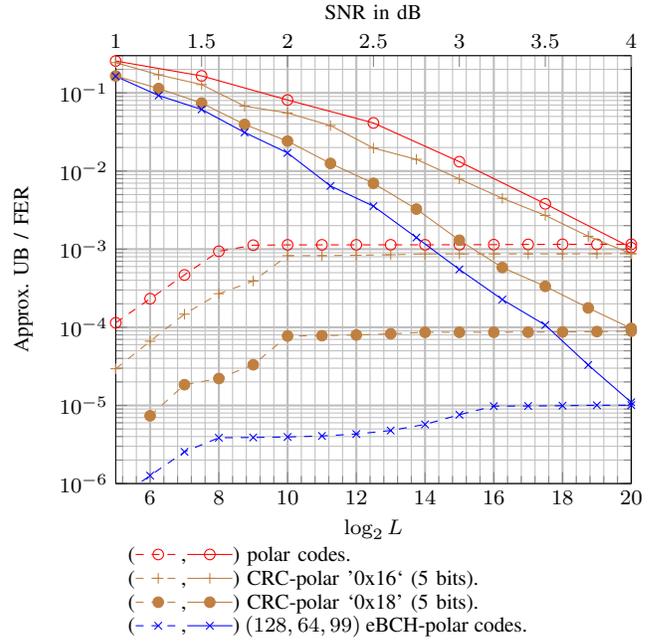
\begin{figure}
	\centering
	\begin{tikzpicture}[scale=1]
	\footnotesize
	\begin{semilogyaxis}[
	ymin=0.000001,
	ymax=0.3,
	xmin = 5,
	xmax = 20,
	minor x tick num=4,
	minor y tick num=5,
	major grid style={thick},
	grid=both,
	axis x line*=bottom,
	xlabel = $\log_2 L$,
	xlabel near ticks,
	ylabel = Approx. UB / FER,
	legend style={at={(0.5,-0.2)},anchor=north}
	]
	
	\addplot[color=red,mark=o,dashed,mark options=solid] coordinates {
		(5,0.00011421)
		(6,0.00023211)
		(7,0.0004679)
		(8,0.00093949)
		(9,0.0011242)
		(10,0.0011345)
		(11,0.0011355)
		(12,0.0011358)
		(13,0.0011363)
		(14,0.0011372)
		(15,0.0011391)
		(16,0.0011429)
		(17,0.0011504)
		(18,0.0011541)
		(19,0.0011541)
		(20,0.0011541)
	};\label{polar1}
	
	\addplot[color=brown,mark=+,dashed,mark options=solid] coordinates {
		(5,2.9474e-05)
		(6,6.6317e-05)
		(7,0.00014737)
		(8,0.00026895)
		(9,0.00039053)
		(10,0.00082562)
		(11,0.0008266)
		(12,0.00083555)
		(13,0.00084874)
		(14,0.00086751)
		(15,0.0008678)
		(16,0.00086793)
		(17,0.00086869)
		(18,0.00086993)
		(19,0.00087268)
		(20,0.00087618)
	};\label{polarcrc161}
	
	\addplot[color=brown,mark=*,dashed,mark options=solid] coordinates {
		(6,7.3686e-06)
		(7,1.8421e-05)
		(8,2.2106e-05)
		(9,3.3158e-05)
		(10,7.739e-05)
		(11,7.8173e-05)
		(12,7.9818e-05)
		(13,8.2628e-05)
		(14,8.6292e-05)
		(15,8.635e-05)
		(16,8.6472e-05)
		(17,8.6704e-05)
		(18,8.7163e-05)
		(19,8.8137e-05)
		(20,8.9280e-05)
	};\label{polarcrc181}
	
	\addplot[color=blue,mark=x,dashed,mark options=solid] coordinates {
		(5,6.2218e-07)
		(6,1.2644e-06)
		(7,2.5489e-06)
		(8,3.8608e-06)
		(9,3.8903e-06)
		(10,3.9493e-06)
		(11,4.0672e-06)
		(12,4.3032e-06)
		(13,4.775e-06)
		(14,5.7187e-06)
		(15,7.6062e-06)
		(16,9.7935e-06)
		(17,9.8382e-06)
		(18,9.9274e-06)
		(19,1.0059e-05)
		(20,1.0086e-05)
	};\label{polarbch1}
	\end{semilogyaxis}
	
	\begin{semilogyaxis}[
	legend style={ at={(0,0)},anchor=south west},
	ymin=0.000001,
	ymax=0.3,
	hide y axis,
	axis x line*=top,
	xlabel near ticks,
	xmin = 1,
	xmax = 4,
	xlabel = SNR in dB,
	ylabel = FER,
	]
	
	\addplot[red, mark = o]
	table[x=snr,y=fer]{snr fer
		1 0.25641
		1.5 0.16529
		2 0.081301
		2.5 0.041322
		3 0.013193
		3.5 0.003811
		4 0.0010478
		4.5 0.00023687
	};\label{polar2}
	
	\addplot[brown, mark = +]
	table[x=snr,y=fer]{snr fer
		1 0.2445
		1.25 0.17065
		1.5 0.12788
		1.75 0.067889
		2 0.055494
		2.25 0.038256
		2.5 0.019701
		2.75 0.014136
		3 0.0079403
		3.25 0.0044899
		3.5 0.0027076
		3.75 0.0014632
		4 0.00087688
		4.25 0.00033601
		4.5 0.0001573
		4.75 6.8e-05
		5 3e-05
	};\label{polarcrc162}
	
	\addplot[brown, mark = *]
	table[x=snr,y=fer]{snr fer
		1 0.16502
		1.25 0.1139
		1.5 0.074239
		1.75 0.039557
		2 0.02419
		2.25 0.012533
		2.5 0.0069745
		2.75 0.0032637
		3 0.0013033
		3.25 0.00058307
		3.5 0.00033484
		3.75 0.00017724
		4 9.7e-05
		4.25 4e-05
		4.5 0
		4.75 0
		5 0
	};\label{polarcrc182}
	
	\addplot[blue, mark = x]
	table[x=snr,y=fer]{snr fer
		1 0.16367
		1.25 0.092678
		1.5 0.061538
		1.75 0.031221
		2 0.017094
		2.25 0.0064508
		2.5 0.0035581
		2.75 0.0014095
		3 0.00055089
		3.25 0.00022698
		3.5 0.00010693
		3.75 3.3e-05
		4 1.1e-05
		4.25 3e-06
		4.5 1e-06
		4.75 0
		5 1e-06
	};\label{polarbch2}
	
	\end{semilogyaxis}
	
	\end{tikzpicture}
	\parbox{0.6\columnwidth}{
		\footnotesize
		(\ref{polar1},\ref{polar2}) polar codes.\\
		(\ref{polarcrc161},\ref{polarcrc162}) CRC-polar ’0x16‘ (5 bits).\\
		(\ref{polarcrc181},\ref{polarcrc182}) CRC-polar ‘0x18’ (5 bits).\\
		(\ref{polarbch1},\ref{polarbch2}) $(128,64,99)$ eBCH-polar codes.
	}
	\caption{ An example of AUBs for different ($128, 64$) codes (optimized for $\SI{4}{dB}$). The solid lines describe the relation between FER and SNR of the codes, while the dashed lines show the AUBs with list size $L$.}
	\label{fig:eg}
\end{figure}

Fig.~\ref{fig:eg} is an AUBs (dashed lines) example for 4 different polar codes (designed and operated at $\SI{4}{dB}$, SCL with $L=32$). The list size is doubled until the AUBs converge. Without convergence, we would only get a lower bound of the UB.  The simulation results (solid lines) show clearly that the ML-performance (at $\SI{4}{dB}$) can be well approximated by the converged AUB. All AUBs shown in our work are based on experiments where the AUBs converged.

\begin{table}
	\caption{Estimated ML- and SC-performance of $(128,64)$ eBCH, RM and polar codes ($\SI{4}{dB}$)}
	\label{tab:dp}
	\centering
	{\renewcommand{\arraystretch}{1.5}
		\begin{tabular}{|c|c|c|c|}				
			\hline
			& Parameter& AUB & Estimated SC FER \\
			\hline
			polar & &  \num{0.0011541} & \num{0.0021507}\\
			\hline
			\multirow{2}{*}{RM-polar} & $d=8$ & \num{0.0011541} & \num{0.0021507} \\
			\cline{2-4} & $d=16$ & \num{1.0887e-05} &  \num{0.022155} \\
			\hline
			\multirow{6}{*}{eBCH-polar} 
			& $k^\prime=106$ & \num{0.0011541} & \num{0.0021507} \\
			\cline{2-4} & $k^\prime=99$ &  \num{1.0086e-05} &\num{0.0067334}\\
			\cline{2-4} & $k^\prime=92$ &  \num{1.0086e-05} &\num{0.0067334}\\
			\cline{2-4} & $k^\prime=85$ & \num{5.3637e-06} &\num{0.00854}  \\
			\cline{2-4} & $k^\prime=78$ & \num{7.445e-06} & \num{0.020806}\\
			\hline
		\end{tabular}
	}
\end{table}	

\begin{table}
	\caption{Best CRC codes for $(128,64)$ CRC-codes (optimized for $\SI{4}{dB}$)}
	\label{tab:crc}
	\centering
	{\renewcommand{\arraystretch}{1.5}
		\begin{tabular}{|c|c|c|c|}	
			\hline
			$\ell_{\rm CRC}$ &Polynomial& AUB & Estimated SC FER\\
			\hline
			3&0x5 &\num{1.9433e-04} &\num{0.004275}\\
			\hline
			4&0xC &\num{1.6791e-04} &\num{0.0051077}\\
			\hline
			5&0x18& \num{8.9280e-05} &\num{0.0064029}\\
			\hline
			6&0x2D & \num{5.8441e-06} &\num{0.0078608}\\
			\hline
			7&0x72& \num{3.2828e-06} &\num{0.0095868}\\
			\hline
			8&0xA6&\num{2.4525e-06}&\num{0.011465} \\
			\hline
		\end{tabular}
	}
\end{table}
Table~\ref{tab:dp} shows the AUB and estimated FER under SC decoding of all options for ($128, 64$) polar, RM-polar and eBCH-polar codes. For CRC-polar codes, the CRC polynomials are optimized for the AUB with exhaustive search. The performance of the $(128,64)$ CRC-polar code with polynomial '0x44' by list size $L=32$ is shown in~\cite{liva2016code}. This code has the AUB \num{3.5926e-06}, which is very close to the best \num{3.2828e-06}.

\begin{remark}
	For the AUB of $(128,64)$ codes at $\SI{4}{dB}$, (\ref{eq:aub}) is not a good approximation because $\SI{4}{dB}$ is not high enough. Therefore, not only $A_{\rm min}$ and ${\rm d}_{\rm min}$ are important. For example, for $(128,64)$ codes at $\SI{4}{dB}$, the CRC-polar code with polynomial '0x72' has ${\rm d}_{\rm min}=12$ and $A_{\rm min}=117$, while the eBCH-polar code with $k^\prime=85$ has ${\rm d}_{\rm min}=16$ and $A_{\rm min}=45592$. However, Table~\ref{tab:dp} and Table~\ref{tab:crc} show that the CRC-polar code has a lower AUB than the eBCH-polar code.
\end{remark}

\section{Design Rules for List Decoding}\label{sec:opt}
In Sec.~\ref{sec:pre}, we introduced three kinds of polar codes with improved distance spectrum. Their SC- and ML-performance could be adjusted via $\ell_{\rm CRC}$, $d$ and $k^\prime$. However, the FER estimation for polar codes with list decoding is not easy. We use three conjectures to simplify analysis: \\
Consider two polar codes $A$ and $B$ with the same code length $n$ and message length $k$. The SC- and ML-performance can be described by ${\rm FER_{SC}}(\cdot)$ and ${\rm FER_{ML}}(\cdot)$.
\begin{conjecture}
	If one of the following conditions is fulfilled, then code $A$ outperforms code $B$ by any list sizes $L\in (1,2^k)$ at high SNR.
	\begin{itemize}
		\item[1.] ${\rm FER_{SC}}(A) \leq {\rm FER_{SC}}(B)$, ${\rm FER_{ML}}(A) < {\rm FER_{ML}}(B)$.
		\item[2.] ${\rm FER_{SC}}(A) < {\rm FER_{SC}}(B)$, ${\rm FER_{ML}}(A) \leq {\rm FER_{ML}}(B)$.
	\end{itemize}
	
	An example is shown in Fig.~\ref{fig:same_dp}. The codes with similar AUB ($\approx10^{-5}$ at $\SI{4}{dB}$) have similar ML-performance. However, the codes with better SC-performance perform better with small list size, i.e., the codes with better SC-performance need smaller list size to achieve the same ML bound with SCL decoding. The dashed curves denote the estimated SC-performance of the codes. Three different codes with similar AUB perform similar for a large list size ($L=128$), while the curves are sorted by the SC-performance for a small list size ($L=4$).
\end{conjecture}
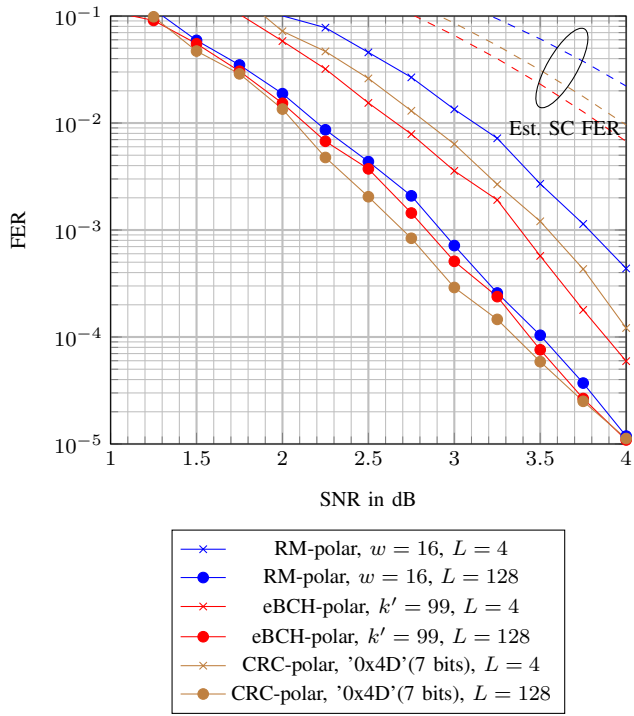
\begin{figure}
	\centering
	\begin{tikzpicture}[scale=1]
	\footnotesize
	\begin{semilogyaxis}[
	legend style={at={(0.5,-0.2)},anchor=north},
	ymin=0.00001,
	ymax=0.1,
	minor x tick num=4,
	minor y tick num=5,
	major grid style={thick},
	grid=both,
	xmin = 1,
	xmax = 4,
	xlabel = SNR in dB,
	ylabel = FER,
	]
	
	\addplot[blue, mark = x]
	table[x=snr,y=fer]{snr fer
		1 0.38023
		1.25 0.29412
		1.5 0.2611
		1.75 0.18282
		2 0.1004
		2.25 0.077882
		2.5 0.045725
		2.75 0.026717
		3 0.01343
		3.25 0.0071984
		3.5 0.0027026
		3.75 0.0011392
		4 0.00043729
	};\addlegendentry{RM-polar, $w=16$, $L=4$}

	\addplot[blue, mark = *]
	table[x=snr,y=fer]{snr fer
		1 0.14881
		1.25 0.11534
		1.5 0.059242
		1.75 0.034892
		2 0.018815
		2.25 0.0086453
		2.5 0.004359
		2.75 0.0020868
		3 0.00071546
		3.25 0.00025786
		3.5 0.00010379
		3.75 3.7094e-05
		4 1.1853e-05
	};\addlegendentry{RM-polar, $w=16$, $L=128$}
	
	\addplot[red, mark = x]
	table[x=snr,y=fer]{snr fer
		1 0.28818
		1.25 0.2008
		1.5 0.16611
		1.75 0.10384
		2 0.058377
		2.25 0.03201
		2.5 0.015415
		2.75 0.0078753
		3 0.003573
		3.25 0.0019107
		3.5 0.00057254
		3.75 0.00017916
		4 5.9474e-05
	};\addlegendentry{eBCH-polar, $k^\prime=99$, $L=4$}
	
	\addplot[red, mark = *]
	table[x=snr,y=fer]{snr fer
		1 0.11038
		1.25 0.090992
		1.5 0.055127
		1.75 0.030497
		2 0.015298
		2.25 0.0067467
		2.5 0.00373
		2.75 0.0014412
		3 0.00050963
		3.25 0.00023834
		3.5 7.5954e-05
		3.75 2.6606e-05
		4 1.09e-05
	};\addlegendentry{eBCH-polar, $k^\prime=99$, $L=128$}
	
	\addplot[brown, mark = x]
	table[x=snr,y=fer]{snr fer
		1 0.31546
		1.25 0.27027
		1.5 0.22523
		1.75 0.15601
		2 0.072098
		2.25 0.046707
		2.5 0.026116
		2.75 0.013009
		3 0.0063557
		3.25 0.0026736
		3.5 0.0012059
		3.75 0.00043149
		4 0.00012137
	};\addlegendentry{CRC-polar, '0x4D'(7 bits), $L=4$}
	
	\addplot[brown, mark = *]
	table[x=snr,y=fer]{snr fer
		1 0.12516
		1.25 0.098425
		1.5 0.047059
		1.75 0.028818
		2 0.013499
		2.25 0.0047653
		2.5 0.002047
		2.75 0.00083728
		3 0.00029026
		3.25 0.0001461
		3.5 5.8783e-05
		3.75 2.505e-05
		4 1.125e-05
	};\addlegendentry{CRC-polar, '0x4D'(7 bits), $L=128$}
	
	\addplot[blue, dashed]
	table[x=snr,y=fer]{snr fer
		1 0.78565
		1.25 0.70908
		1.5 0.62232
		1.75 0.5296
		2 0.436
		2.25 0.34655
		2.5 0.2655
		2.75 0.19573
		3 0.13861
		3.25 0.09409
		3.5 0.061061
		3.75 0.037758
		4 0.022155
	};
	\addplot[red, dashed]
	table[x=snr,y=fer]{snr fer
		1 0.70509
		1.25 0.60838
		1.5 0.50516
		1.75 0.40247
		2 0.30703
		2.25 0.22395
		2.5 0.15605
		2.75 0.10381
		3 0.065907
		3.25 0.039928
		3.5 0.023086
		3.75 0.012748
		4 0.0067334
	};
	\addplot[brown, dashed]
	table[x=snr,y=fer]{snr fer
		1 0.78946
		1.25 0.69756
		1.5 0.59252
		1.75 0.48207
		2 0.37481
		2.25 0.27814
		2.5 0.19695
		2.75 0.1331
		3 0.085894
		3.25 0.052977
		3.5 0.031259
		3.75 0.01767
		4 0.0095868
	};
	\end{semilogyaxis}
	\draw[rotate around={60:(6,5)}] (6, 5) ellipse (.6 and 0.2);
	\node[text width=3cm] at (6.8,4.2) 
	{Est. SC FER};
	\end{tikzpicture}
	\caption{Three $(128,64)$ polar codes with same AUB, optimized for $\SI{4}{dB}$}
	\label{fig:same_dp}
\end{figure}
\begin{conjecture}
	If code $A$ has better SC-performance and worse ML-performance, i.e., if
	\begin{itemize}
		\item[] ${\rm FER_{SC}}(A) < {\rm FER_{SC}}(B)$, ${\rm FER_{ML}}(A) > {\rm FER_{ML}}(B)$
	\end{itemize}
	then there is a list size $L^\prime$ with the following property. Code $A$ outperforms code $B$ for list size $L$ where $L<L^\prime$, while code $B$ performs better for $L>L^\prime$ at high SNR. An example for $(128,64)$ codes is shown in Fig.~\ref{fig:exap_c2}.
\end{conjecture}
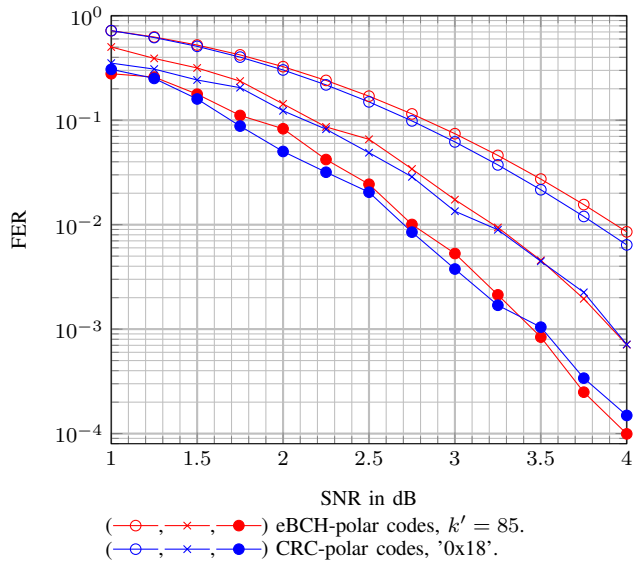
\begin{figure}
	\centering
	\begin{tikzpicture}[scale=1]
	\footnotesize
	\begin{semilogyaxis}[
	legend style={at={(0.5,-0.2)},anchor=north},
	ymin=0.00008,
	ymax=1,
	minor x tick num=4,
	minor y tick num=5,
	major grid style={thick},
	grid=both,
	xmin = 1,
	xmax = 4,
	xlabel = SNR in dB,
	ylabel = FER,
	]
	\addplot[red, mark = o]
	table[x=snr,y=fer]{snr fer
		1 0.72049
		1.25 0.6263
		1.5 0.52457
		1.75 0.42207
		2 0.32558
		2.25 0.24046
		2.5 0.16992
		2.75 0.11488
		3 0.074313
		3.25 0.046027
		3.5 0.027323
		3.75 0.015572
		4 0.00854
	};\label{bch1}
	\addplot[red, mark = x]
	table[x=snr,y=fer]{snr fer
		1 0.50251
		1.25 0.39062
		1.5 0.31646
		1.75 0.23697
		2 0.14327
		2.25 0.086281
		2.5 0.06566
		2.75 0.034141
		3 0.017361
		3.25 0.0093967
		3.5 0.0045523
		3.75 0.0019527
		4 0.00071019
	};\label{bch2}
	\addplot[red, mark = *]
	table[x=snr,y=fer]{snr fer
		1 0.27855
		1.25 0.25907
		1.5 0.17762
		1.75 0.11111
		2 0.08285
		2.25 0.042034
		2.5 0.02439
		2.75 0.010046
		3 0.0052927
		3.25 0.0021288
		3.5 0.0008385
		3.75 0.0002492
		4 9.9364e-05
	};\label{bch4}
	
	\addplot[blue, mark = o]
	table[x=snr,y=fer]{snr fer
		1 0.7202
		1.25 0.61861
		1.5 0.50972
		1.75 0.40189
		2 0.3028
		2.25 0.21794
		2.5 0.14987
		2.75 0.098533
		3 0.061989
		3.25 0.037355
		3.5 0.021589
		3.75 0.011985
		4 0.0064029
	};\label{crc1}
	\addplot[blue, mark = x]
	table[x=snr,y=fer]{snr fer
		1 0.35211
		1.25 0.30864
		1.5 0.24272
		1.75 0.20492
		2 0.12315
		2.25 0.081967
		2.5 0.048876
		2.75 0.028719
		3 0.013448
		3.25 0.008935
		3.5 0.0044607
		3.75 0.0022504
		4 0.00071343
		4.25 0
		4.5 0
		4.75 0
		5 0
	};\label{crc2}
	\addplot[blue, mark = *]
	table[x=snr,y=fer]{snr fer
		1 0.30675
		1.25 0.25126
		1.5 0.15974
		1.75 0.087873
		2 0.050201
		2.25 0.031786
		2.5 0.020458
		2.75 0.0084818
		3 0.0037619
		3.25 0.0016923
		3.5 0.0010445
		3.75 0.00033935
		4 0.00014919
		4.25 0
		4.5 0
		4.75 0
		5 0
	};\label{crc4}
	
	\end{semilogyaxis}
	\end{tikzpicture}
	\parbox{0.65\columnwidth}{
		\footnotesize
		(\ref{bch1},\ref{bch2},\ref{bch4}) eBCH-polar codes, $k^\prime=85$.\\
		(\ref{crc1},\ref{crc2},\ref{crc4}) CRC-polar codes, '0x18'.
	}
	\caption{An example for conjecture 2, optimized for $\SI{4}{dB}$, $L=\{1,2,4\}$}
	\label{fig:exap_c2}
\end{figure} 
\begin{conjecture}
	By list decoding with fixed $L$, the SC-performance of polar codes becomes more important for larger code dimension $k$ and vice versa, the SC-performance becomes less important for smaller $k$. 
\end{conjecture}
Now we use the conjectures to find the best eBCH codes. Fig.~\ref{fig:opt_128_64} shows that $(128,64,99)$ and $(128,64,85)$ codes perform similar for list size 32, $(4096,2048,\geq 3915)$ and $(4096,2048, 3903)$ codes perform similar for list size 2. Using Conjecture 2, we know for $L\leq L^\prime =32$, the $(128,64,99)$ eBCH-polar code should be used, and for $L\geq L^\prime =2$, the $(4096,2048, 3903)$ eBCH-polar code performs better. From Table~\ref{tab:dp}, we know that the $(128,64,99)$ eBCH-polar code is the most polar-like code among the $(128,64,k^\prime)$ codes, that improve the minimum distance. This result could be extended by using Conjecture 3: For polar codes with dimension $64\leq k\leq 2048$ and decoding list size $2\leq L\leq 32$, we should use the most polar-like codes that improve the minimum distance. Some simulation results are shown in Fig.~\ref{fig:bch_opt_1024} and Fig.~\ref{fig:bch_opt_256} for eBCH-polar codes with $n=\{256,1024\}, {\rm code \ rate}=\{1/4,1/2,3/4\}$ and decoding list size 8.

\begin{figure}
	\centering
	\begin{tikzpicture}[scale=1]
	\footnotesize
	\begin{semilogyaxis}[
	legend style={at={(0.4,-0.2)},anchor=north},
	ymin=0.000001,
	ymax=0.1,
	minor x tick num=4,
	minor y tick num=5,
	major grid style={thick},
	grid=both,
	xmin = 1,
	xmax = 4.25,
	xlabel = SNR in dB,
	ylabel = FER,
	]
	\addplot[red, mark = o]
	table[x=snr,y=fer]{snr fer fer_ml
		1 0.26178 0.26178
		1.25 0.14903 0.14754
		1.5 0.12315 0.12315
		1.75 0.072993 0.072263
		2 0.058343 0.058343
		2.25 0.037092 0.037092
		2.5 0.02076 0.02076
		2.75 0.014006 0.014006
		3 0.0079491 0.0079491
		3.25 0.0052681 0.0052681
		3.5 0.003187 0.003187
		3.75 0.0017844 0.0017844
		4 0.0011568 0.0011568
		4.25 0.00055737 0.00055737
		4.5 0.00026633 0.00026633
		4.75 0.00015707 0.00015707
		5 6.0819e-05 6.0819e-05
	};\addlegendentry{$(128,64,\geq 106)$ eBCH-polar codes, $L=32$, ${\rm d}_{\rm min}=8$}
	
	\addplot[brown, mark = o]
	table[x=snr,y=fer]{snr fer fer_ml
		1 0.16639 0.12978
		1.25 0.091912 0.079044
		1.5 0.065833 0.046083
		1.75 0.028885 0.023108
		2 0.019131 0.015688
		2.25 0.0084746 0.0072034
		2.5 0.0036889 0.0028774
		2.75 0.0015116 0.0012093
		3 0.00062233 0.00052898
		3.25 0.00027078 0.00023017
		3.5 9.4532e-05 8.697e-05
		3.75 2.9116e-05 2.7369e-05
		4 1.0336e-05 1.0233e-05
		4.25 2.9e-06 2.9e-06
		4.5 1.2e-06 1.2e-06
	};\addlegendentry{$(128,64,99)$ eBCH-polar codes, $L=32$, ${\rm d}_{\rm min}=12$}
	\addplot[blue, mark = o]
	table[x=snr,y=fer]{snr fer fer_ml
		1 0.15432 0.10031
		1.25 0.11862 0.073547
		1.5 0.073746 0.037611
		1.75 0.042088 0.025673
		2 0.018925 0.01022
		2.25 0.0080574 0.0047538
		2.5 0.0040553 0.0023521
		2.75 0.0015964 0.00094186
		3 0.0005292 0.00033869
		3.25 0.00020386 0.00013659
		3.5 7.6255e-05 5.2616e-05
		3.75 2.5894e-05 1.4548e-05
		4 9e-06 4.7e-06
		4.25 2.5e-06 1.4e-06
		
	};\addlegendentry{$(128,64,85)$ eBCH-polar codes, $L=32$, ${\rm d}_{\rm min}=16$}
	
	\addplot[red, mark = *, dashed]
	table[x=snr,y=fer]{snr fer fer_ml
		1 0.68493 0
		1.25 0.34722 0
		1.5 0.067249 0.00067249
		1.75 0.015154 0
		2 0.0019016 5.7049e-05
		2.25 0.00015305 0
		2.5 1.3288e-05 3.0865e-07
		2.75 1.4e-06 0
	};\addlegendentry{$(4096,2048,\geq 3915)$ eBCH-polar codes, $L=2$, ${\rm d}_{\rm min}=32$}
	\addplot[blue, mark = *, dashed]
	table[x=snr,y=fer]{snr fer fer_ml
		1 0.72464 0
		1.25 0.40486 0
		1.5 0.08058 0
		1.75 0.018426 0
		2 0.0017684 0
		2.25 0.00020814 0
		2.5 1.5876e-05 0
		2.75 1.4e-06 0
		
	};\addlegendentry{$(4096,2048, 3903)$ eBCH-polar codes, $L=2$, ${\rm d}_{\rm min}=48$}	
	
	\end{semilogyaxis}
	\end{tikzpicture}
	\caption{eBCH-polar codes by list size $L^\prime$, optimized for $\{4, 2.5\}$ dB}
	\label{fig:opt_128_64}
\end{figure}

\begin{figure}
	\centering
	\begin{tikzpicture}[scale=1]
	\footnotesize
	\begin{semilogyaxis}[
	legend style={at={(0.5,-0.2)},anchor=north},
	ymin=0.000001,
	ymax=1,
	minor x tick num=4,
	minor y tick num=5,
	major grid style={thick},
	grid=both,
	xmin = -3,
	xmax = 6,
	xlabel = SNR in dB,
	ylabel = FER,
	]
	
	\addplot[blue, mark = o]
	table[x=snr,y=fer]{snr fer
		-3 0.3876
		-2.75 0.20121
		-2.5 0.13717
		-2.25 0.056625
		-2 0.025113
		-1.75 0.009552
		-1.5 0.0052538
		-1.25 0.0022661
		-1 0.00081072
		-0.75 0.00054644
		-0.5 0.00020849
		-0.25 8.4064e-05
		0 3.6222e-05
		0.25 1.3813e-05
		0.5 0
	};\addlegendentry{$(1024,256,\geq 873)$, ${\rm d}_{\rm min}=32$ (original)}
	
	\addplot[blue, mark = *]
	table[x=snr,y=fer]{snr fer
		-3 0.44444
		-2.75 0.28902
		-2.5 0.13755
		-2.25 0.067843
		-2 0.024402
		-1.75 0.0073589
		-1.5 0.0015025
		-1.25 0.00031349
		-1 5.7079e-05
		-0.75 9.7e-06
		-0.5 0
		-0.25 0
		0 0
		0.25 0
		0.5 0
	};\addlegendentry{$(1024,256, 863)$, ${\rm d}_{\rm min}=48$}
	
	\addplot[blue, mark = x]
	table[x=snr,y=fer]{snr fer
		-3 0.4902
		-2.75 0.35336
		-2.5 0.16584
		-2.25 0.096061
		-2 0.034928
		-1.75 0.0085448
		-1.5 0.0021295
		-1.25 0.00048833
		-1 7.5527e-05
		-0.75 1.3752e-05
		-0.5 0
		-0.25 0
		0 0
		0.25 0
		0.5 0
	};\addlegendentry{$(1024,256, 783)$, ${\rm d}_{\rm min}=64$}

	\addplot[red, mark = o]
	table[x=snr,y=fer]{snr fer
		0 0.96154
		0.25 0.97087
		0.5 0.7874
		0.75 0.46083
		1 0.29674
		1.25 0.13736
		1.5 0.039047
		1.75 0.012149
		2 0.0045009
		2.25 0.0017388
		2.5 0.00065012
		2.75 0.00024692
		3 0.0001077
		3.25 4.1853e-05
		3.5 1.4791e-05
	};\addlegendentry{$(1024,512,\geq 953)$, ${\rm d}_{\rm min}=16$ (original)}
	
	\addplot[red, mark = *]
	table[x=snr,y=fer]{snr fer
		0 0.97087
		0.25 0.91743
		0.5 0.73529
		0.75 0.49751
		1 0.27397
		1.25 0.14306
		1.5 0.044307
		1.75 0.0093336
		2 0.0019252
		2.25 0.00031641
		2.5 3.9333e-05
		2.75 3e-06
		3 0
		3.25 0
		3.5 0
	};\addlegendentry{$(1024,512, 943)$, ${\rm d}_{\rm min}=24$}
	
	\addplot[red, mark = x]
	table[x=snr,y=fer]{snr fer
		0 0.9901
		0.25 0.98039
		0.5 0.84034
		0.75 0.56818
		1 0.3876
		1.25 0.16367
		1.5 0.05005
		1.75 0.017112
		2 0.003897
		2.25 0.0007111
		2.5 6.6544e-05
		2.75 7.5e-06
		3 0
		3.25 0
		3.5 0
	};\addlegendentry{$(1024,512, 903)$, ${\rm d}_{\rm min}=32$}
	
	\addplot[brown, mark = o]
	table[x=snr,y=fer]{snr fer
		3 0.9901
		3.25 0.97087
		3.5 0.74074
		3.75 0.57143
		4 0.27701
		4.25 0.12739
		4.5 0.034095
		4.75 0.010481
		5 0.0022666
		5.25 0.00062954
		5.5 0.00028838
		5.75 0.00011617
		6 4.4872e-05
		6.25 1.9078e-05
		6.5 0
	};\addlegendentry{$(1024,768,\geq 993)$, ${\rm d}_{\rm min}=8$ (original)}
	
	\addplot[brown, mark = *]
	table[x=snr,y=fer]{snr fer
		3 0.98039
		3.25 0.93458
		3.5 0.8547
		3.75 0.66225
		4 0.33898
		4.25 0.15385
		4.5 0.045434
		4.75 0.0075626
		5 0.00129
		5.25 0.00019366
		5.5 9.2e-06
		5.75 0
		6 0
		6.25 0
		6.5 0
	};\addlegendentry{$(1024,768,983)$, ${\rm d}_{\rm min}=12$}
	
	\addplot[brown, mark = x]
	table[x=snr,y=fer]{snr fer
		3 0.9901
		3.25 0.97087
		3.5 0.8547
		3.75 0.67114
		4 0.33898
		4.25 0.18622
		4.5 0.04386
		4.75 0.01271
		5 0.0018966
		5.25 0.00024216
		5.5 1.5691e-05
		5.75 0
		6 0
		6.25 0
		6.5 0
	};\addlegendentry{$(1024,768,963)$, ${\rm d}_{\rm min}=16$}
	
	\end{semilogyaxis}
	\end{tikzpicture}
	\caption{$(1024,k,k^\prime)$ eBCH-polar codes with $L=8$, optimized for $\{-0.25,3,6\}$ dB}
	\label{fig:bch_opt_1024}
\end{figure}

\begin{figure}
	\centering
	\begin{tikzpicture}[scale=1]
	\footnotesize
	\begin{semilogyaxis}[
	legend style={at={(0.5,-0.2)},anchor=north},
	ymin=0.000001,
	ymax=1,
	minor x tick num=4,
	minor y tick num=5,
	major grid style={thick},
	grid=both,
	xmin = -3,
	xmax = 7,
	xlabel = SNR in dB,
	ylabel = FER,
	]
	\addplot[blue, mark = o]
	table[x=snr,y=fer]{snr fer
		-3 0.28409
		-2.75 0.22321
		-2.5 0.12376
		-2.25 0.08673
		-2 0.068259
		-1.75 0.042517
		-1.5 0.02588
		-1.25 0.014102
		-1 0.0063387
		-0.75 0.0046788
		-0.5 0.0019711
		-0.25 0.0013032
		0 0.00064776
		0.25 0.00031252
		0.5 0.00020311
		0.75 0.00010468
		1 5.3134e-05
	};\addlegendentry{$(256,64,\geq 199)$, ${\rm d}_{\rm min}=16$ (original)}
	
	\addplot[blue, mark = *]
	table[x=snr,y=fer]{snr fer
		-3 0.35211
		-2.75 0.23697
		-2.5 0.14903
		-2.25 0.085179
		-2 0.047962
		-1.75 0.028106
		-1.5 0.017437
		-1.25 0.0075609
		-1 0.005478
		-0.75 0.0015564
		-0.5 0.0007332
		-0.25 0.00026897
		0 0.00011049
		0.25 3.2853e-05
		0.5 7.5e-06
		
	};\addlegendentry{$(256,64,191)$, ${\rm d}_{\rm min}=32$}
	
	\addplot[blue, mark = x]
	table[x=snr,y=fer]{snr fer
		-3 0.4902
		-2.75 0.39683
		-2.5 0.31056
		-2.25 0.25974
		-2 0.12953
		-1.75 0.087336
		-1.5 0.05152
		-1.25 0.024131
		-1 0.012686
		-0.75 0.005316
		-0.5 0.002165
		-0.25 0.00076363
		0 0.00023286
		0.25 6.0803e-05
		0.5 1.5076e-05
		0.75 0
		1 0
	};\addlegendentry{$(256,64,127)$, ${\rm d}_{\rm min}=48$}

	\addplot[red, mark = o]
	table[x=snr,y=fer]{snr fer
		0.5 0.47847
		0.75 0.36496
		1 0.23256
		1.25 0.16367
		1.5 0.081169
		1.75 0.040584
		2 0.015509
		2.25 0.0078431
		2.5 0.0031707
		2.75 0.0014178
		3 0.00058586
		3.25 0.00018041
		3.5 7.0617e-05
		3.75 2.6814e-05
		4 6.5e-06
		4.25 2.1e-06
	};\addlegendentry{$(256,128,\geq 199)$, ${\rm d}_{\rm min}=16$ (original)}
	
	\addplot[red, mark = *]
	table[x=snr,y=fer]{snr fer
		0.5 0.67568
		0.75 0.4878
		1 0.34014
		1.25 0.21739
		1.5 0.14368
		1.75 0.07722
		2 0.036778
		2.25 0.016534
		2.5 0.0058207
		2.75 0.0016325
		3 0.00057241
		3.25 0.0001334
		3.5 2.4845e-05
		3.75 5e-06
		
	};\addlegendentry{$(256,128,191)$, ${\rm d}_{\rm min}=24$}
	
	\addplot[red, mark = x]
	table[x=snr,y=fer]{snr fer
		0.5 0.81967
		0.75 0.70922
		1 0.59172
		1.25 0.51813
		1.5 0.34483
		1.75 0.23753
		2 0.13055
		2.25 0.064935
		2.5 0.035689
		2.75 0.01411
		3 0.0045249
		3.25 0.0011383
		3.5 0.00028905
		3.75 7.3257e-05
		4 1.186e-05
		4.25 2.6e-06
	};\addlegendentry{$(256,128,159)$, ${\rm d}_{\rm min}=28$}

	\addplot[brown, mark = o]
	table[x=snr,y=fer]{snr fer
		3 0.73529
		3.25 0.5848
		3.5 0.46948
		3.75 0.32258
		4 0.23474
		4.25 0.13055
		4.5 0.075075
		4.75 0.036456
		5 0.012222
		5.25 0.0066881
		5.5 0.003892
		5.75 0.0015612
		6 0.00059667
		6.25 0.0001951
		6.5 6.0982e-05
		6.75 2.2473e-05
		7 8e-06
	};\addlegendentry{$(256,192,\geq 231)$, ${\rm d}_{\rm min}=8$ (original)}
	
	\addplot[brown, mark = *]
	table[x=snr,y=fer]{snr fer
		3 0.86207
		3.25 0.71942
		3.5 0.66225
		3.75 0.52083
		4 0.31646
		4.25 0.15848
		4.5 0.10428
		4.75 0.054675
		5 0.022578
		5.25 0.0071762
		5.5 0.0019959
		5.75 0.0006829
		6 0.0001305
		6.25 3.2998e-05
		6.5 4.1e-06
		
	};\addlegendentry{$(256,192,223)$, ${\rm d}_{\rm min}=12$}
	
	\addplot[brown, mark = x]
	table[x=snr,y=fer]{snr fer
		3 0.9901
		3.25 0.92593
		3.5 0.86957
		3.75 0.76923
		4 0.60606
		4.25 0.4878
		4.5 0.28409
		4.75 0.18657
		5 0.095694
		5.25 0.047801
		5.5 0.019936
		5.75 0.0070053
		6 0.001873
		6.25 0.00043961
		6.5 8.1095e-05
		6.75 1.9488e-05
		7 2.6e-06
	};\addlegendentry{$(256,192,207)$, ${\rm d}_{\rm min}=14$}
	
	\end{semilogyaxis}
	\end{tikzpicture}
	\caption{$(256,k,k^\prime)$ eBCH-polar codes with $L=8$, optimized for $\{0.5,3.5,6.5\}$ dB}
	\label{fig:bch_opt_256}
\end{figure}

\section{Low Weight Bits Construction }\label{sec:df}
\begin{figure}
	\centering
	\begin{tikzpicture}[scale=1]
	\footnotesize
	\begin{semilogyaxis}[
	legend style={at={(0,0)},anchor=south west},
	ymin=0.000001,
	ymax=0.1,
	minor x tick num=4,
	minor y tick num=5,
	major grid style={thick},
	grid=both,
	xmin = 1.5,
	xmax = 4.25,
	xlabel = SNR in dB,
	ylabel = FER,
	]
	
	\addplot[red, mark = o]
	table[x=snr,y=fer]{snr fer
		1 0.2439
		1.25 0.16077
		1.5 0.10204
		1.75 0.051813
		2 0.03701
		2.25 0.013951
		2.5 0.0066613
		2.75 0.0035786
		3 0.0019118
		3.25 0.00067025
		3.5 0.00017353
		3.75 5.6549e-05
		4 1.95e-05
		4.25 5.4092e-06
		4.5 1.2284e-06
	};\addlegendentry{eBCH $k^\prime = 99$, $L=8$}
	
	\addplot[blue, mark = o]
	table[x=snr,y=fer]{snr fer
		1 0.21834
		1.25 0.17301
		1.5 0.089445
		1.75 0.070323
		2 0.043066
		2.25 0.012752
		2.5 0.0070452
		2.75 0.0028932
		3 0.0016396
		3.25 0.00054568
		3.5 0.00015209
		3.75 4.3015e-05
		4 1.2943e-05
		4.25 3.5834e-06
	};\addlegendentry{$N_{\rm df}=7$, $L=8$}
	
	\addplot[red, mark = x]
	table[x=snr,y=fer]{snr fer fer_ml
		1 0.16639 0.12978
		1.25 0.091912 0.079044
		1.5 0.065833 0.046083
		1.75 0.028885 0.023108
		2 0.019131 0.015688
		2.25 0.0084746 0.0072034
		2.5 0.0036889 0.0028774
		2.75 0.0015116 0.0012093
		3 0.00062233 0.00052898
		3.25 0.00027078 0.00023017
		3.5 9.4532e-05 8.697e-05
		3.75 2.9116e-05 2.7369e-05
		4 1.0336e-05 1.0233e-05
		4.25 2.9e-06 2.9e-06
		4.5 1.2e-06 1.2e-06
	};\addlegendentry{eBCH $k^\prime = 99$, $L=32$}
	
	\addplot[blue, mark = x]
	table[x=snr,y=fer]{snr fer fer_ml
		1 0.1642 0.12315
		1.25 0.097561 0.08
		1.5 0.06689 0.050167
		1.75 0.033979 0.025824
		2 0.013795 0.0093806
		2.25 0.0087966 0.0075651
		2.5 0.0033163 0.0025867
		2.75 0.0015948 0.0012439
		3 0.00057873 0.0004572
		3.25 0.00020068 0.00017459
		3.5 7.6572e-05 7.1212e-05
		3.75 2.298e-05 2.0222e-05
		4 6.9e-06 6.8e-06
		4.25 1.9e-06 1.9e-06
	};\addlegendentry{$N_{\rm df}=7$, $L=32$}
	
	\end{semilogyaxis}
	\end{tikzpicture}
	\caption{Comparison between eBCH-polar code and our construction, optimized for $\SI{4}{dB}$}
	\label{fig:new_dp}
\end{figure}
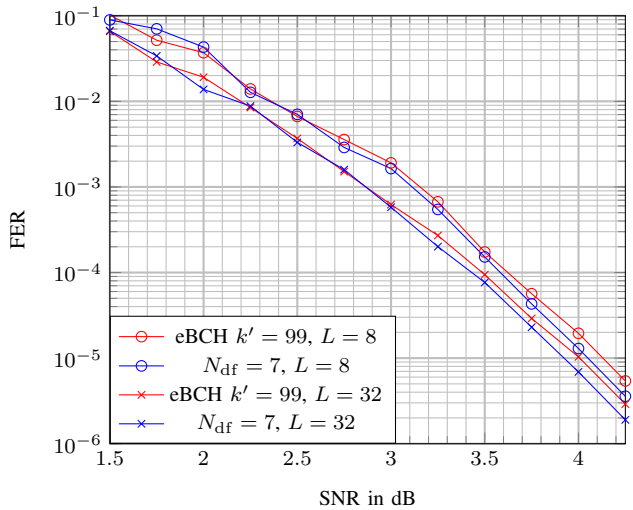

We propose a simple construction of polar codes with improved distance spectrum via dynamically frozen bits.
\begin{itemize}
	\item[1.] Design an $(n,k+N_{\rm df})$ polar code.
	\item[2.] Find the set $\mathcal{I}_{\rm min}$ of information bits according to low row weights in $\mathbb{F}^{\otimes\log n}$.
	\item[3.] Add $N_{\rm df}$ linearly independent constraints among the bits in $\mathcal{I}_{\rm min}$.
\end{itemize}
The parameter $N_{\rm df}$ denotes the number of dynamically frozen bits and describes how many reliable bits (under SC decoding) are sacrificed.
\begin{table}
	\caption{Estimated ML- and SC-performance of the $(128,64)$ polar codes with dynamically frozen bits (optimized for $\SI{4}{dB}$)}
	\label{tab:df}
	\centering
	{\renewcommand{\arraystretch}{1.5}
		\begin{tabular}{|c|c|c|}	
			\hline
			& AUB & Estimated SC FER\\
			\hline
			$N_{\rm df}=7$ &7.1666e-06 &0.0067334\\
			\hline
			eBCH-polar, $k^\prime=99$ &1.0086e-05 &0.0067334\\
			\hline
		\end{tabular}
	}
\end{table}
Table~\ref{tab:df} shows the SC-performance and the AUB of polar codes ($N_{\rm df}=7$) with constraints:
\begin{equation}
\begin{split}
&u_{85}=u_{99}=u_{113}=u_{57}\oplus u_{83},\\
&u_{89}=u_{101}=u_{57},\\
&u_{98}=u_{105}=u_{83}.
\end{split}
\end{equation}
Using Conjecture 1, the polar codes with $N_{\rm df}=7$ outperform the $(128,64,99)$ eBCH-polar code for any list sizes. Fig.~\ref{fig:new_dp} shows the comparison between eBCH-polar codes and our scheme by decoding list size 8 and 32. The gain is $\SI{0.1}{dB}$ at FER $10^{-5}$.

\section{Conclusions}\label{sec:con}
In this work, we review state-of-the-art polar code constructions for distance improvement and analyze their distance spectrum. A heuristic construction is proposed to optimize the list decoding performance for a certain range of dimension $k$ and list size $L$. In addition, a polar code construction based on dynamically frozen bits and "low weight bits" is proposed, which provides better performance than eBCH-polar codes.

\bibliographystyle{IEEEtran}
\bibliography{IEEEabrv,references}

\end{document}